# The Most Common Habitable Planets II - Salty Oceans in Low Mass Habitable Planets and Global Climate Evolution

R. Pinotti [1,2] * and G. F. Porto de Mello[1]

*1 – Observatório do Valongo, Universidade Federal do Rio de Janeiro - UFRJ, Ladeira Pedro Antônio 43, 20080-090, Rio de Janeiro, RJ, Brasil*
*2 - Petrobras, Av. Henrique Valadares, 28, 20231-030, Rio de Janeiro, RJ, Brazil*

**ABSTRACT**

Global climate evolution models for habitable earthlike planets do not consider the effect of ocean salinity on land ice formation through the hydrological cycle. We consider two categories of such planets: planets with deep oceans, but intrinsically high salinities due to the weaker salt removal process by hydrothermal vents; and planets with shallow oceans, where the increase in salt content and decrease in ocean area during the onset of glaciation cause a negative feedback, helping delay the spread of land ice. We developed a toy climate model of a habitable planet on the verge of an ice age, using a range of initial salt concentrations. Planets with deep oceans and high salinity show considerable increase in the time necessary to fill arctic land with ice sheets, up to 23% considering the maximum salinity range. For planets with shallow oceans, the effect of intrinsic high salinity is reinforced by the negative feedback, counteracting positive feedbacks like the ice-albedo and Croll-Milankovitch perturbations, to the point of effectively terminating land ice sheet growth rate during the simulated timescale. We also apply this model to the putative ocean of early Mars, finding intermediate results: salinity probably did not play a role in the evolution of Mars´ climate, considering the timescale of its ice ages. We conclude that this phenomenon is essentially an abiotic self-regulation mechanism against ice ages and should be regarded in the context of habitable planets smaller and drier than the Earth, which may well represent the bulk of habitable planets.

## 1 INTRODUCTION

Climate evolution studies of habitable planets have undergone two different waves of stimulus over the past century: the first one in the wake of the exploration of Venus and Mars by robotic probes, concretely starting the field of comparative planetology in the 1960s (e. g., Kiefer et al. 1992; Barrie 2008; Launius 2012). The second one followed the discovery of the first extrasolar planet around a main sequence star (Mayor & Queloz, 1995), almost a quarter of a century ago. The field has also received recent and strong input from studies of the climate of our own planet, with important and practical results such as long-term alterations in climate patterns due to the increasing anthropogenic greenhouse effect.

Nowadays the number of confirmed discovered extrasolar planets has surpassed 4,200 (Schneider et al. 2011), and, with the aid of fast computing, we are able to simulate the climate of not just hypothetical, but also actual planets, constrained by data on radiation flux from the star, size, and density of the planet, and, in some cases, data on atmospheric composition (Wallack et al. 2019; Molaverdikhani et al. 2020). However, most of the data available on atmospheric composition still refer to uninhabitable planets like Hot Jupiters, but observatories like the James Webb Space Telescope, the E-ELT and TESS may change this scenario in the near future, enabling the analysis of the atmospheres of as yet undiscovered nearby rocky planets in the habitable zone of their stars. This tantalizing possibility was further fueled by the discovery of a possibly habitable rocky planet in the habitable zone of Proxima Centauri, our closest neighbor in the Galaxy (Anglada-Escudé et al. 2016; Ribas et al.2016; Turbet et al. 2016). Unfortunately, the measurement of key variables such as the water content of such planets remain beyond the current technological capabilities and must be assumed in simulations.

*email: rpinotti@astro.ufrj.br

The presence of water in the liquid state at planetary surfaces has been elected as a prime criterion to define their habitability, helping to frame the so called orthodox stellar habitable zone around main sequence stars of different masses (Kasting et al. 1993; Franck et al. 2000; Kopparapu et al. 2013). The origin of the Earth's water and its total content are still under debate, as well as the possibility that planets similar to our own in the habitable zones of other stars may be as rich, or even richer in water than the Earth (Noack et al. 2016). On the other hand, details about water delivery to planets inside the habitable zone as well as volatile removal by high-energy emissions may play a crucial role on planetary water content. Particularly for very low mass stars, total or near total removal of planetary volatiles by prolonged exposure to XUV radiation and particle winds may result in the average habitable planet, at least at the low end of the planetary mass range, being considerably water poor as compared to the Earth (Lammer et al. 2009; Ribas et al. 2016; Bolmont et al. 2017; Dong et al. 2017; Airapetian et al. 2017), even though models suggest that some degree of water retention, contingent upon initial water inventory, remains a clear possibility (Bolmont et al. 2017). Suggestions that most of the potentially habitable rocky planets are probably small as well as water deficient (Lissauer 2007; Pinotti 2013) strengthen the importance of detailed studies of such worlds, including climate models and conditions under which they might hold on to their volatile inventories and remain habitable.

If water deficient habitable planets are as numerous as these studies suggest, then climate models would benefit by taking into account the total superficial water volume as a parameter, as well as its distribution along the planet´s area. We will see later that this distinction is quite relevant.

Moreover, smaller planets would be expected to have oceans with higher salt content, since the main salt removal process, namely the ocean ridge hydrothermal circulation system (Langmuir & Broecker 2012), would be less intense, along with tectonic activity in general, due to the faster cooling of such planets. This reasoning, coupled to the fact that water evaporation and freezing are affected by salt content, leads to the possibility that salt content may play an important role in the climate evolution of such planets. More specifically, the decrease of water evaporation would affect the formation of glaciers on land, since the balance of water content in the atmosphere depends on it, and the depression of freezing point would delay the formation of sea ice. Moreover, for planets with shallow oceans, the surface seawater volume is smaller still as compared to planets with oceans with average water depth on the order of kilometers, and thus the decline of ocean volume during glaciation might be sufficient to increase salinity over time, creating a negative-feedback effect on glaciation. Although the relation between salinity and water freezing point has been taken into account in previous studies (Cullum, Stevens & Joshi 2016; del Genio et al. 2019, Olson 2020), the effect of salinity on evaporation and land glacier formation has not, as hasn´t either the negative-feedback effect on glaciation for shallow oceans.

In order to study the effects of ocean salinity on the climate of habitable planets with varying degrees of surface water extension and volume, we developed a toy climate model, with 26 variables and 26 algebraic-differential equations, and investigated the salinity-evaporation relation for a specific climate dynamics, namely, a planet on the verge of an ice age.

The paper is organized as follows: in section 2 we discuss the possible "pristine" surface water content of exoplanets; section 3 contains a discussion about the loss of volatiles caused by stellar wind and radiation; in section 4 we discuss the variables that determine the salt content on the oceans of the Earth and of exoplanets; in section 5 there is a description of our climate toy model; section 6 shows the results of the simulations and a discussion about them, including a customized simulation for the planet Mars; finally, section 7 presents our conclusions.

**2 SURFACE WATER VOLUME IN ROCKY PLANETS**

Since we will perform simulations of the climate of planets of different sizes, an *a priori* step is an assessment of their possible surface water volume. The problem of the origin of Earth's oceans is still unresolved, and researchers have arguments for both wet-endogenous and exogenous sources (van Dishoeck et al. 2014; D'Angelo et al. 2019; Ida, Yamamura & Okuzumi 2019). The wet-endogenous theory links the water to solids that formed the Earth, and outgassed to the surface (Elkins-Tanton 2012; Lebrun et al. 2013; Noack et al. 2016). The exogenous one calls for water delivery by small bodies or by nebular gas (Genda and Ikoma 2008; Morbidelli et al. 2012; Raymond et al. 2014), coming from outer regions of the solar system, after the formation of the Earth and generally considered as a late addition. Constraints like the protoplanetary disk temperature in the region where Earth is believed to have been formed, and the deuterium to hydrogen ratio of the water in the Earth preclude a simple answer, and a combination of both sources is a possibility. A recent claim (Piani et al. 2020) adds to the controversy by claiming that enstatite chondrite meteorites match the isotopic composition of terrestrial rocks and may be representative of the material that formed the Earth. This new result attributes to them a sufficiently high hydrogen content to account for at least three times the mass of water in Earth´s oceans.

The amount of water delivered to rocky extrasolar planets by comets and asteroids is probably a stochastic process, which depends not only on the protoplanetary disk composition, number of planetesimals, and the position of the snowline (Ciesla et al. 2015), but also on the number, size and radial distribution of other planets in the systems, variables with strong stochastic nature, given the available sample of extrasolar multiplanetary systems and planet formation simulations.

In what follows, we assume the wet-endogenous hypothesis to hold to a first approximation, since it allows simple deterministic correlations to be developed. Let us represent the total surface water volume of a general planet, by Eq. (1):

$$V_s = h\theta 4\pi R_p^2 \quad (1)$$

where $V_s$ is the total volume, $h$ is the average water depth of the oceans, $\theta$ is the initial fraction of the planetary surface covered by water, and $R_p$ is the planetary radius. Then, the ratio between the surface water volume and the total surface area of the planet is given by $V_s/A_p = h\theta$, where $A_p = 4\pi R_p^2$ stands for the total planetary surface area. In our case it is important not to use a single term to express the water volume to total surface ratio, since the evaporation process depends on the water area exposed to the atmosphere, individually expressed by $\theta$.

Our premise that the main source of water **is** endogenous allows us to write a relation between this volume and the planet's mass as Eq. (2):

$$V_s = \delta \frac{\rho_p}{\rho_w} \frac{4}{3} \pi R_p^3 \quad (2)$$

where $\rho_p$ is the average planetary density, $\rho_w$ is the water density, and $\delta$ is the fraction of water mass compared to the total mass of the planet. Then, combining Eqs. (1) and Eq. (2), we obtain Eq. (3):

$$h\theta = \delta \frac{\rho_p}{\rho_w} \frac{1}{3} \pi R_p \quad (3)$$

Average planetary density differs from uncompressed density (the latter being a better indicator of planetary composition). Our goal is to model planets less massive than the Earth, and the uncompressed density of Mars (lying at the low end of the mass range of interest) is 96% of its mean density. Even Earth and Venus have uncompressed densities amounting to 80% of their mean densities, thus this fact introduces but a small inaccuracy in our model.

This equation indicates a linear relation between planetary radius and the product $h\theta$, considering $\delta$ and $\rho_p$ as relatively constant for rocky planets in the Habitable Zone. However, we must bear in mind that many factors may turn this assumption wrong, due to different planetary compositions, including percentage of radionuclides, which affect volcanism and plate tectonics, and consequently outgassing efficiency. In addition, Eq (3) is probably an *upper limit* of water coverage, since water loss processes, mainly due to stellar activity, have an increasing importance for smaller planets**.** Lastly, Cowan & Abbot (2014) modelled the partition of water between the surface and the mantle in terrestrial planets, following degassing and subduction, concluding that planets more massive than the Earth are more capable to hold water in the mantle (a fact of consequence to the habitability of superearths). This effect is correspondingly much less relevant to undermassive planets, which are our focus, so we hold Eq. (3) as correct to a first approximation. Volatile loss for undermassive planets is probably a much more relevant process as compared to Earth-sized planets and superearths. We also note that the water content in Earth´s mantle remains controversial within nearly two orders of magnitude (Cowan & Abbot 2014).

According to Eq (3), for sufficiently high values of R, $\theta_0$ tends to 1, and we will have ocean planets, as described in theoretical studies (Adams, Seager & Elkins-Tanton 2008; Kaltenegger & Sasselov 2013; Kitzmann et al. 2015; Simpson 2019,). However, in our work we will focus on low mass habitable planets, which tend to have both less extensive and shallower oceans.

Although the wet-endogenous hypothesis leads to simple correlations between planet size and water coverage, the parameter $\delta$ will generally be unknown, and $\rho_p$ will be accessible only for planets discovered by the transit method and followed-up with radial velocity observations. While we expect the fraction of these to constantly increase, even so density will probably be known with considerable uncertainty at least until current methods improve by a large factor. This restricts our ability to make reliable predictions. The exogenous hypothesis, while stochastic in nature, offers the possibility of a wider range of surface water content on exoplanets. It is currently uncertain whether the most probable mechanism accounting for Earth´s water inventory is either entirely endogenous (Piani et al. 2020) or at least partly exogenous (Raymond & Izidoro 2017). It seems likely that real planets will sport a wide range of water delivery and migration histories, as well as formation conditions and water cycling. Our toy model does not depend on the origin of the surface water in exoplanets, its only premise being that planets smaller than the Earth tend to be drier, due mainly to considerations described in section 3. Our assumptions can thus be accommodated to a varying range of sources for a planet's water inventory: resulting models can correspondingly be tweaked to adjust for a considerable range of initial conditions. We limit ourselves in the present exploratory analysis to relations that be analytically derived. Finally, we note that the amount itself of exogenous sources of water is expected to depend on the size of the planet, given the Hill sphere.

**3 PARENT STAR AND LOSS OF VOLATILES**

The development of the previous section gives us a direction on how dry a rocky planet may become as a function of its size. Using the Earth as the basis for extrapolation, with around 70% of its surface covered by water, and considering its average ocean depth of 3.7 km a constant, we derive, for planets with radii 75% and 50% that of the Earth's, a water coverage of 52.5% and 35% respectively.

However, this is but a rough estimate of the initial water coverage at the time a planet is born. Stellar winds, X-ray and UV (XUV) radiation may erode the atmosphere and volatiles from the surface of rocky planets. This process is dependent on many factors involving both planetary and stellar properties, and a large variety of end results concerning volatile retention is expected. For solar-type stars in general, with masses between 0.6 and 1.5 solar masses, the high-activity phase is very short-lived and directly connected to the evolution of rotation (Wood et al. 2002, 2005; Ribas et al. 2005; Ribas et al. 2010; do Nascimento et al. 2016) for both high-energy XUV photons and winds, and only very low mass planets are affected, as is the case with Mars. For lower mass stars inside the red dwarf regime, a much longer phase is maintained (West et al. 2008) starting around type M3. Connected to the much closer-in habitable zones for such less luminous stars, this fact makes for much higher volatile losses expected for rocky planets inside the habitable zones of stars less massive than about ~0.5 solar masses (Luger & Barnes 2015). Even more severe conditions are expected for the less massive rocky planets (our focus in this work), since their lower gravity facilitates atmosphere loss and atmospheric replenishment from volcanism is expected to be shorter lived.

Possible exceptions to this general scenario include water outgassing from the mantle (Moore & Cowan 2020). Sequestering water in the mantle with posterior degassing might replenish surface water content, provided an Earth-like deep-water cycle can be kept on and sufficient water was originally sequestered within the mantle. Again, out focus lies in modeling the lesser massive planets, and these should be both originally endowed with less water and able to transition to a stagnant lid regime as exemplified by Mars. Such transition would possibly stop both degassing and regassing (Schaefer & Sasselov 2015), preserving the water inventories in the surface and mantle reservoirs and sharply slowing down their water exchange. Thus, the possibility of water replenishing from the mantle is less relevant for planets substantially less massive than the Earth. We note, though, that such degassing would involve juvenile water which would contribute to lower oceanic salt content, being a potential counter to the effect we describe. In the following we assume that, while this remains a possibility, degassing should not greatly modify our modelling.

Yet another factor hinders the retention of volatiles for less massive planets: the core dynamo which generates a protective magnetosphere is expected to be weaker from the start and subside faster than those of bigger planets. Zuluaga et al. (2013) modelled the evolution of planetary magnetospheric properties for a range of assumptions about mass, chemical composition, and thermal evolution, and calculated the magnetic standoff radius, which is an underestimate of the actual size of the dayside magnetosphere. They find that even planets with ~0.5 Earth masses are able to achieve and maintain sizable standoff magnetospheric radii, not much unlike planets as massive as the Earth, as long as they remain tidally unlocked. These results hold even during the star's initial phase of strong winds. For tidally locked planets, much diminished standoff radii are found, yet still some measure of protection is to be found. Tidally locked rotation, or capture into low order rotational resonances, for planets inside the habitable zone is essentially ubiquitous for stars less massive than about 0.7 solar masses (Porto de Mello et al. 2006). Turbet el al. (2016) find that the onset of low-order resonances considerably improves volatile retention, at least for planets as massive as the Earth, and, also, that there is little difference in climate stabilization between the 2:1 and 3:2 resonances, provided that the planet remains asynchronous. Such resonances should not be uncommon.

Additionally, Leconte et al. (2015) report that thermal tides, even in relatively thin atmospheres, can drive the rotation of an Earthlike planet away from synchronicity and achieve stable asynchronous equilibrium spin states. Such states are possible if thermal tide amplitudes exceed a threshold, which can plausibly be met for planets with at least 1 bar atmospheres around stars with spectral types earlier than about M0, corresponding to masses larger than about 0.50 solar masses. Thinner atmospheres in planets orbiting less massive stars probably are much more prone to synchronous or nearly synchronous rotational states.

Notwithstanding these possibilities, considerably larger volatile loss from various physical mechanisms are expected for rocky planets orbiting stars less massive than about 0.6-0.7 solar masses, increasing in severity as we consider less massive planets.

Considering these complications, and recognizing that actual planets probably span a much wider range of properties, we finally assume that the Earth has suffered minor volatile loss, and we opted to perform simulations for three different planetary radii relative to Earth's (100%, 75% and 50%). Accordingly, the values of $\theta_0$ were chosen as 70%, 40% and 10% respectively, which incorporates a non-linear reduction factor on the initial values estimated form Eq. (3). Our assumed values fall in the range between so-called aqua planets and land planets (Kodama et al. 2019), a distinction with some importance for the definition of the inner edge of Habitable Zones.

**4 SOURCES AND SINKS OF OCEAN SALT**

We shall consider in our model that, for lower planetary radius, plate tectonics will probably play a shorter and less intense role in many aspects of planetary climate, including its role as a sink of oceanic salt, through hydrothermal circulation. Although the phenomenon of plate tectonics depends on a number of variables (Stern et al. 2018; Stamenković & Breuer 2014; Karato 2014; Noack et al 2014), including the abundance of radionuclides (Frank et al. 2014), our focus is on planets that have significantly cooled off (age on the order of a few Gyr).

This reasoning, added to the probable smaller total water volume indicated by Eq (3), leads to the conclusion that the ocean of these worlds will probably be richer in salt, for the input of salt into the oceans through rivers is linked to the hydrologic cycle, which in turn is a process relatively independent from plate tectonics in the time-scales considered in this work ($10^2$ - $10^5$ years).

The above arguments lead us to expect that not only would the volume of surface water change with planetary radius, but also its composition. It was only recently, with the discovery of the ocean ridge hydrothermal circulation (Langmuir and Broecker 2012; Hovland, Rueslåtten & Johnsen 2018) that the enigma of seawater composition was solved. The continuous input of salt from rivers to the oceans must be balanced by a sink, otherwise the salt content of the oceans would have reached the saturation level in a geologically short time. Table 1 shows the compositions of some elements in rain, rivers, seawater and hydrothermal fluid. The hydrothermal systems, which remove sodium and chlorine, among other elements, and is a source of others, process the entire volume of the oceans in tens of millions of years, and is directly related to the existence of plate tectonics. In planets smaller than the Earth, plate tectonics is expected to be less intense – even absent in some cases – after some Gyr, compared to that of the Earth, leading to the conclusion that the ocean ridge hydrothermal circulation would probably be less efficient in removing salt from the oceans. The oceans, in turn, would still be receiving the flow of rivers due to the hydrological cycle, so that their salt content would be higher than the Earth's oceans.

| Element | Rain | Rivers | Seawater | Hydrothermal Fluid |
|---------|------|--------|----------|--------------------|
| Ca | 0.65 | 13.3 | 412 | 1200 |
| Mg | 0.14 | 3.1 | 1290 | 0 |
| Na | 0.56 | 5.3 | 10770 | - |
| K | 0.11 | 1.5 | 380 | 975 |
| Sulphate | 2.2 | 8.9 | 2.688 | 28 |
| Cl | 0.57 | 6 | 19000 | - |
| Si | 0.3 | 4.5 | 2 | 504 |
| Fe | 0 | 0.03 | 0.002 | 168 |
| Mn | 0 | 0.007 | 0.0002 | 41 |
| Flux to oceans (kg/yr) | | $4 \times 10^{16}$ | | $3 \times 10^{13}$ |

Table 1 – Composition of Earth's water (Langmuir and Broecker 2012) – concentrations in parts per million

Due to many other factors influencing plate tectonics, it is not possible to directly correlate planetary size with salt content in the oceans, so we will assume, in our simulations, a range of salt content for each planetary radius considered, between the value of Earth's oceans (35,000 ppm) and the value of near saturation (260,000 ppm).

**5 THE MODEL**

In order to assess the influence of ocean salinity on the evaporation rate and ice formation in oceans and on land, we developed a toy model for a planet with roughly the same characteristics of the Earth. The model performs a dynamic water mass balance between oceans, the atmosphere and land ice, starting with a state that goes in the direction of an ice age, that is, one in which ice formation rate on land is positive and unbounded.

Since the majority of rainwater on Earth falls on the oceans or is returned to the oceans through rivers, the dynamic water budget of the atmosphere will consider only the evaporation from oceans, multiplied by a factor that corrects for water not returned to the oceans, and the formation of glaciers on land. The average cloud coverage of the planet is assumed constant, since its modeling is known as quite a difficult task, and a complete model is beyond the objective of this study. Moreover, the cloud coverage is not supposed to vary significantly during the time scale of the simulations.

The global surface energy balance will be given by Eq. (4), the well-known equilibrium equation between the radiative energy flux emitted by the entire planetary surface and the absorbed stellar flux:

$$(1 - A) F \pi R^2 = 4 \varepsilon \pi \sigma R^2 T^4 \quad (4)$$

The average planetary albedo, $A$, will be a function of time in the simulations, as well as the area covered by ice, sea and land, each of which having a specific albedo. $T$ is the average temperature of the upper troposphere (usually known as skin temperature), whose heat exchange mechanism is mainly radiative, $F$ is the stellar bolometric radiative flux, $\sigma$ is the Stefan-Boltzmann constant, $\varepsilon$ is the emissivity, which is close to 0.9 in the infrared, and $R$ is the radius of the planet.

Eq. (4) assumes that the term $dT/dt$, due to the accumulation of thermal energy in the atmosphere (Pinotti 2013), is negligible, that is, the variation of $A$ over time is small enough, so that thermal equilibrium in the atmosphere is always attained and the variation of $T$ over time is given by an algebraic relation with $A$. And, in fact, the time required for the ice/sea/land coverage to change appreciably is large compared with the time step for the integration of the ensemble of differential equations. Eq. (4) also assumes that the planet is a fast rotator, so that there is no significant temperature difference between the nightside and the dayside.

The model planet was set with non-zero inclination of the rotation axis relative to the ecliptic, similarly to the Earth, and we divide it into three distinct climatic regions: tropical, temperate, and arctic. Let also the inclination be the same, 23.44°, or $0.1302\pi$ rad. Assuming that the average stellar flux intercepted by the planet is the one dictated by an equinox configuration (seasonal changes are not our focus), we determine the values of intercepted radiation flux for each region. The results are shown in Table 2.

Note that the three regions encompass the northern and southern portions, for both are identical in terms of radiative balance in the equinox configuration, and consequently are considered as one in the simulations. Therefore, when a region is mentioned hereafter, we mean both the northern and southern ones added together.

| Region | Area intercepting stellar radiation | Total area | Fraction of Total area |
|--------|-------------------------------------|------------|------------------------|
| Tropical | $1.5478\ R^2$ | $4.9980\ R^2$ | 0.3977 |
| Temperate | $1.5055\ R^2$ | $6.5318\ R^2$ | 0.5198 |
| Arctic | $0.0883\ R^2$ | $1.0367\ R^2$ | 0.0825 |
| Total for the planet | $\pi R^2$ | $4\pi R^2$ | 1 |

Table 2 – Areas for the climatic regions of a planet of radius R, with inclination of rotation axis of 23.4369° relative to the ecliptic, and at the equinox point of its orbit.

Eq. (4) will be used in a modified form to estimate the temperature of the upper troposphere of the three regions ($T_A$, $T_{Te}$ and $T_{Tr}$ for the arctic, temperate and tropical zones, respectively), that is, the intercepting and total areas will not be $\pi R^2$ and $4\pi R^2$ respectively, but the values for each region given in Table 1.

In order to calculate $T_A$, $T_{Te}$ and $T_{Tr}$, using the data on Table 1, we also need values for the albedos of the three regions. We assume a range of initial conditions: their exact values do not affect the simulation results in a significant way and simply aim at providing boundary conditions for launching the simulations:

- each region will have its own ocean, with albedo given by Table 3, and the same initial water coverage fraction ($\theta_0$) which defines the simulation case; that is, the main simulation cases are defined by $\theta_0 = 70\%$, $\theta_0 = 40\%$ and $\theta_0 = 10\%$; this water coverage area will be a variable in the simulations ($\theta_i = \theta_i(t)$, with $I = 1, 2, 3$), since evaporation will affect the ocean volumes, and to a different degree;

- the initial average depth of the oceans is 3,700 m, similar to that of the Earth's; this value, nearing Earth´s condition, has been chosen as a baseline; in the section dedicated to shallow oceans we use the value of 500 m; a further decrease to the level of tens of meters place the environment away from an oceanic domain, and the probable net of water bodies, separated by small land irregularities, is better described as a lake system, being outside the scope of this work;
- the bare land, with albedo given by Table 3, initially covers the remaining fraction ($1- \theta_0$); this coverage will also change over time, as the seas shrink due to evaporation;
- the tropical region is free of sea or land ice in the model, since its surface temperature is always above the freezing point of water;
- the temperate region has an initial ice coverage of 1% over land (fraction of total temperate area = 0.01 (1- $\theta_0$)) and 1% over sea (fraction of total temperate area = 0.01 $\theta_0$); both ice coverages will be time variables in the model; this assumption is based on the probability that Earth-like planets with oceans will also have high mountains in temperate regions where permanent glaciers reside, and where there might be some sea ice during part of the year;
- the arctic region has an initial ice coverage of 15% over land (fraction of total arctic area = 0.15*(1- $\theta_0$)) and 15% over sea (fraction of total arctic area = 0.15 $\theta_0$); both ice coverages will be time variables in the model; these values are compatible with the arctic region having average surface temperature well below 0 ºC, as shown below;
- the whole planet has a fixed cloud coverage of 30%, evenly distributed over sea and land; this value is below that of present day Earth, because we will be dealing with a planet entering an ice age, with a drier atmosphere.

| Surface constitution | Albedo |
|---|---|
| Sea water | 0.06 |
| Land | 0.27 |
| Ice | 0.35 |
| Cloud | 0.50 |

Table 3 – Albedos for each surface constitution: typical values are adopted, considering the range of possibilities.

Under these assumptions, applying Eq. (4) separately to each climatic zone, we calculate the initial albedos and upper troposphere temperatures for each region, and for each value of initial $\theta_0$, which defines each simulation case. The results are shown in Table 4 and Table 5.

The significant temperature differences between the regions is due to the assumption that they are independent of each other. On Earth, the differences are dampened by air and ocean currents. But there will be no sea currents in our model, since the seas are isolated; air currents are indirectly assumed (this topic will be discussed below), but the air has less heat capacity than water, so the model of three isolated regions is probably not far from the reality of a planet with the characteristics as we describe.

| Region | $\theta_0 = 0.7$ | $\theta_0 = 0.4$ | $\theta_0 = 0.1$ |
|---|---|---|---|
| Tropical | 0.2361 | 0.2802 | 0.3243 |
| Temperate | 0.2377 | 0.2813 | 0.3250 |
| Arctic | 0.2599 | 0.2974 | 0.3349 |

Table 4 – Initial albedos for each region and each value of $\theta_0$.

| Region | $\theta_0 = 0.7$ | $\theta_0 = 0.4$ | $\theta_0 = 0.1$ |
|---|---|---|---|
| Tropical | 281.9 | 277.7 | 273.4 |
| Temperate | 261.7 | 257.9 | 253.8 |
| Arctic | 202.5 | 199.9 | 197.2 |

Table 5 – Initial upper troposphere temperatures (K) for each region and each $\theta_0$.

The next step is the estimation of the initial temperature of the lower troposphere, which is different from that of the upper troposphere due to the greenhouse effect and air currents. We will assume that the increment, due to a greenhouse effect, is given by $\psi$, so that $T_{A(S)} = T_A + \psi$, $T_{Te(S)} = T_{Te} + \psi$, and $T_{Tr(S)} = T_{Tr} + \psi$, where $T_{A(S)}$, $T_{Te(S)}$ and $T_{Tr(S)}$ are the surface temperatures (lower troposphere) in the arctic, temperate and tropical regions respectively. The greenhouse effect is a function of the amount of greenhouse gases in the atmosphere. Since $H_2O$ is a strong greenhouse gas, we will link $\psi$ to the relative humidity of the atmosphere, $\phi$, ($\psi = \psi(\phi)$), which will also be a time variable ($\phi = \phi(t)$). We will consider that other greenhouse gases will not change appreciably during the time span of the simulations, so that $\psi = c + f(\phi)$. The term $f(\phi)$ will be linear with respect to $\phi$, that is, $\psi = c + b\phi$, where $b$ and $c$ are constants. This is an admittedly simplistic proxy to the more complex relation between relative humidity and the greenhouse effect, but, then again, our toy model does not presume to be a rigorous one as a whole, and uses simplifications of phenomena and reduction of dimensions, in order to probe a specific relation, that is, the effect of ocean salinity on evaporation (this one modeled with rigorous equations), and, as a consequence, on glaciation. We did not include any variation of $CO_2$ content, for its dynamics during the onset of ice ages on the Earth is similar to that of temperature and humidity, and its greenhouse effect can be viewed as accommodated in our linear relationship with humidity. Moreover, intrinsic $CO_2$ variability has time-scales higher than the ones studied in this work

We will assume that $\phi$ will be the same for all the three regions, that is, we are implicitly assuming that the air currents will transport humidity through the regions and level off any difference. This simplification carries an accuracy cost with it, but since the objective of the toy model is to probe the correlation between the evaporation of salt water and the formation of ice sheets on land, and that the main water reservoir lies in the oceans, rather than in the atmosphere, we believe that the results will not be affected significantly.

Finally, the value of $c$ will be different for each region, thus accounting for air currents, so that the difference between regions becomes less steep and more realistic. Table 6 summarizes the initial values for the temperature of the lower troposphere, for each region. The temperatures of the lower troposphere will be considered as the surface temperatures of the planet.

The tropical and temperate regions will have initial surface temperatures above 0 ºC, and will provide the atmosphere with water vapor through ocean evaporation, whereas the arctic region will have, on the other hand, temperature well below 0 ºC, working essentially as a cold trap.

| Region | $\theta_0 = 0.7$ | $\theta_0 = 0.4$ | $\theta_0 = 0.1$ |
|---|---|---|---|
| Tropical | 308.9 | 304.7 | 300.4 |
| Temperate | 291.7 | 287.9 | 283.8 |
| Arctic | 242.5 | 239.9 | 237.2 |

Table 6 – Initial surface (lower troposphere) temperatures (K) for each region, for each value of $\theta_0$.

The next step, once the initial conditions for the simulations are set, is to model the dynamic water exchange between oceans, the atmosphere and land ice sheets that advance without constraints in our scenario. We consider that the source of formation of sea ice is the freezing of sea water only.

The evaporation rate is known to be directly proportional to the difference between the water vapor pressure of the air in contact with water and the water vapor pressure of the air at a given height (Harbeck 1955; Perry 1997; Babkin 2019). If we also assume that the air in contact with the liquid is saturated, then we can write the following equation for the evaporation rate $H$, in column height per time:

$$H = \alpha\ (p^* - p) \quad (5)$$

where $\alpha$ is a constant, $p^*$ is the water vapor saturation pressure, and $p$ is the water vapor pressure, both at the surface temperature. The water vapor saturation pressure is a function of the temperature only, and we use the well-established Buck equation (Buck 1981):

$$p^* = \exp\left[\left(18.678 - \frac{T}{234.5}\right)\left(\frac{T}{(257.14+T)}\right)\right] \quad (6)$$

where $p^*$ given in kPa, and $T$ in degrees Celsius. On the other hand, $p$ is related to the relative humidity by $\phi = p/p^*$, so that Eq. (5) can be rewritten as

$$H = \alpha\ p^*(1 - \phi) \quad (7)$$

In order to estimate the value of α, we will assume an evaporation rate of the order of $10^3$ mm year$^{-1}$ for sea water at 20 ºC, a value compatible with the average evaporation rate of the oceans of the Earth (Babkin 2019). If we also assume an initial average relative humidity for the toy model as 60% ($\phi_0 = 0.6$), then α = 1,300 mm year$^{-1}$ kPa$^{-1}$. We will discuss later the effect of the choice of α on the results of the simulations.

Here we can appreciate the possible negative feedback on climate by the sea salinity, for the water vapor saturation pressure decreases with an increase in salinity, affecting directly the evaporation rate, and limiting the vapor supply (in fact the only supply) to the atmosphere. As the seas evaporate, the salinity of the water increases, and the evaporation rate decreases (for a fixed value of $\phi$). Over time, as more vapor is fixed in the cold traps of the polar regions and in the land and sea ices of the temperate regions, ϕ will tend to get lower, and the evaporation rate will tend to get higher again, but the changes in albedo (land and land ice will advance over sea - the ice-albedo positive feedback), will force surface temperatures down, which will also affect water vapor saturation pressure. The simulations will indicate the net effect of this negative feedback over time.

Next, we need a quantitative relation between the salinity of water and the depletion of the water vapor saturation pressure $p^*$. We assumed a linear relation between the logarithm of salinity and the logarithm of the relative vapor pressure depletion, using as data the depletion in sea water and in the waters of the Great Salt Lake; the resulting corrected (depleted) water vapor saturation pressure is

$$p' = p^*\left[1 - 10^{((\log(salinity) - \beta)/\gamma)}\right] \quad (8)$$

where salinity is given in parts per million in weight (ppm), $\beta = 5.980$ and $\gamma = 0.831$. This equation agrees very well with established correlations used in industry (Sharqawy et al. 2010; Nayar et al. 2016).

The total water evaporation rate from the oceans, in volume per time (given by column height per time times area), will be:

$$Q = \sum [\alpha p_i'(1 - \phi)A_i] \quad (9)$$

where $i$ represents the climatic region (i = 1, 2, 3) and $A_i$ is the surface area of each of the three oceans ($A_i = A_i\ (\theta_i(t))$). The values of p will evolve with the surface temperature $T_{surf}$, that is, $p_i' = p_i'\ (T_{surf\ i}(t))$. At this point it is interesting to note that there is a positive feedback involved between the vapor pressure depletion with increased salinity and the consequent decrease in evaporation rate. As the vapor supply to the atmosphere is limited by increased salinity, the greenhouse effect caused by water vapor will tend to be lower ($\phi$ will be lower), lowering in its turn the surface temperatures. Also, as the oceans lose water and recede, the area occupied by them will diminish, exposing more bare land, which has a higher albedo, and consequently lowering the surface temperature by means of the radiative balance discussed above. The relation between the volume of each ocean and its evaporation rate is straightforward:

$$\frac{dV_i}{dt} = -\alpha\ p_i'(1 - \phi)A_i \quad (10)$$

However, the dynamic relation between the surface area of the ocean and its volume is an incognita, since we do not know the particulars of the topography of the ocean bed of our putative exoplanets, in spite of the fact that theoretical work has been done to model planetary topography (Landais et al. 2019). The initial ocean surface area and its volume has been calculated based on an average ocean depth of $h_0 = 3.7$ km, an Earth-like value, and the corresponding value of $\theta_0$. Considering that shallow areas become exposed more rapidly than the deeper ones as the oceans evaporate, we adopted a generic relation between the surface area and the volume using an exponential, that is,

$$A_i = A_{i0} \exp\left[-c\left(V_{i0}/V_i(t)\right)\right] \quad (11)$$

For small volume variations, which is the case of the scenarios of deep oceans, the exponential term can be approximated as a linear term with $V_{i0}/V_i(t)$.

The water mass balance between the oceans, the ice glaciers on land and the atmosphere will define the evolution of the relative humidity ($\phi(t)$), that is,

$$\frac{dM}{dt} = [\rho_1 z Q(t) - \rho_2 j G(t)] \quad (12)$$

where $G(t)$ is the land ice area formation rate, $\rho_1$ is the density of water, $\rho_2$ is the density of ice, $j$ is the average height of land glaciers, and $z$ is the fraction of evaporated water that does not return to the oceans in the form of rain, and rivers. On Earth the value of $z$ is around 0.004 (Henshaw, Charlson & Burges 2000), and we chose the value of 0.0035 for our simulations. The term $M(t)$ represents the water mass in the atmosphere, which depends on the volume of the atmosphere and its temperature, which itself is a function of time (through the radiative balance) and defines the water vapor partial pressure $p$, which in turn is related to the relative humidity by $\phi = p/p^*$. Knowledge of the value of $p$ plus the volume of the atmosphere $V$ allows the calculation (through the ideal gas law $pV = nRT$) of the total water mass in the atmosphere, and its derivative, which can be rewritten as $dM/dt = f(\phi(t))$.

The atmosphere we consider will be the lower troposphere, with its temperature dictated by the radiative balance of the upper troposphere plus the greenhouse effect. Since each climatic region will have its own distinct lower troposphere temperature, but the same humidity, the water vapor mass in the atmosphere will be the sum of the mass of the three regions. The total volume of the low troposphere, formed by incondensable gases and water vapor, is given by the area of the planetary surface times an "effective height". We estimated its value by assuming that, with the initial simulation conditions for $\theta_0 = 0.7$ (and $R = R_{Earth}$), the total water vapor in the lower troposphere coincides with the water vapor in the atmosphere of the Earth ($\sim 1.3 \times 10^4$ km$^3$, Henshaw, Charlson & Burges 2000). The value has the same order of magnitude of the troposphere of the Earth. For the smaller planets with $R = 0.75\ R_{Earth}$ and $R = 0.5\ R_{Earth}$ we applied a non-linear reduction factor for the effective height of 0.1 and 0.01 respectively, in order to account for the erosion of the atmosphere by stellar activity, and bearing in mind that Mars' atmosphere is around 99% thinner than the Earth's at the surface of the planet.

Planets of different sizes with the same initial conditions should in principle present the same dynamic behavior of intensive variables (salinity, ice and water coverage as fractions of total planet area), since the relative areas are independent of planet size. The non-linear variation of atmospheric thickness (and its associated greenhouse effect) with radius would in principle disrupt this similarity. However, our simulations indicated that the results, for altering only the atmospheric thickness, are practically the same, and that is because the water mass holdup in the atmosphere is very small compared to the fluxes involved. For example, on Earth the turnover time for water in the atmosphere is only 8 days. Therefore, the results of the simulations, using as the main parameter $\theta_0$, can be interpreted as valid for planets of different sizes. This is also why we used three values of salinity for each $\theta_0$, since there probably are exceptions to our assumption that smaller means drier. Small planets with a higher ocean coverage, which developed high ocean salinities due to a lack and/or weakness of plate tectonics and its associated hydrothermal systems, might well exist.

The land ice formation rate is defined as being proportional to the difference between the surface temperature and the freezing point ($T_f = 273.15$ K). Also, the availability of water vapor in the atmosphere for ice formation must be considered, but since the particulars of weather are beyond the scope of this work, we used an exponential law, so that $G(t)$ was modeled as:

$$G(t) = Y(273.15 - T)\exp(-x/\phi(t)) \quad (13)$$

where $Y$ and $x$ are constants, with values compatible with a growth rate similar to that observed on the onset of ice ages on Earth ($10^3$ years), and on planets with $\theta_0 = 0.7$. $T$ refers to the surface temperature. The sea ice evolution does not alter the water budget calculated in the simulations, since its thickness is small and it is formed mainly from sea water. However, sea ice area does influence the albedo of the arctic regions, and consequently the surface temperature through the radiative balance. The sea ice formation rate used in the simulations is proportional to the difference between the surface temperature and the freezing point of water:

$$\frac{dA_{ice}}{dt} = \Delta(T_f - T_w) \quad (14)$$

where $\Delta$ is a constant, $T_f$ is the freezing point and $T_w$ is the water temperature, which equals the surface temperature. The freezing point of water suffers a depression due to the colligative effect by dissolved salt, according to

$$T_f = 0 - K_f b i \quad (\text{in } °C) \quad (15)$$

where $K_f$ is the cryoscopic constant (for water the value is 1.853 °C kg/mol), $b$ is the molality of the solution, and $i$ is the van't Hoff factor. For NaCl the factor is 2. Hence, for more saline oceans, the rate of sea ice formation will decrease due to the decrease of $T_f$.

The resulting ensemble of 26 algebraic-differential equations and variables was translated to Matlab and integrated by the DASSLC package (Secchi 2010). The simulations are stopped when the sea and land in the arctic regions become fully covered with sea ice and land ice sheets, respectively.

**6 RESULTS AND DISCUSSION**

The results of the simulations are shown separately for the cases of deep oceans (initial average depth of 3.7 km - section 6.1), shallow oceans (initial average depth of 0.5 km - section 6.2) and for a hypothesized ancient Martian ocean (section 6.3).

**6.1 Deep oceans**

Figures 1 and 2 show the results for the most Earth-like scenario of $\theta_0 = 0.7$. Figure 1 shows the evolution of the fraction of arctic ocean area covered by ice, for the three salinities selected, and Figure 2

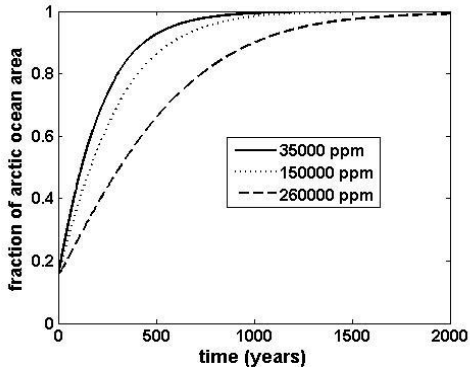

Figure 1 – Evolution of the fraction of arctic ocean area covered by ice, for the case $\theta_0 = 0.7$, average ocean depth 3.7 km, and 3 different ocean salinities

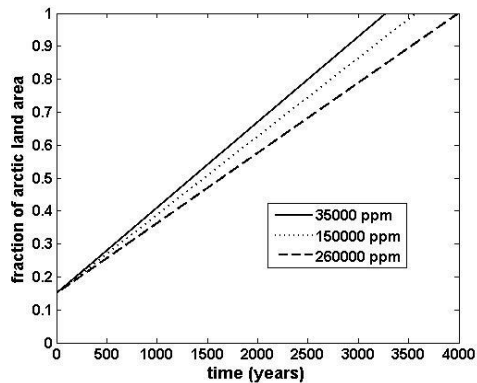

Figure 2 – Evolution of the fraction of arctic land area covered by ice glaciers, for the case $\theta_0 = 0.7$, average ocean depth 3.7 km, and 3 different ocean salinities

shows the evolution of the fraction of arctic land area covered by ice sheets, for the same set of salinities.

The effect of freezing point depression, due to increased water salinity, on the necessary time to freeze completely the arctic oceans is evident; for a salinity of 260,000 ppm the time-scale almost doubles, compared to a planet with an ocean salinity of 35,000 ppm. In the case of the advancing ice sheets on land, the effect of the oceans' salinity is not as large, but there is a significant increase of 23% in the necessary time for them to cover the entire available land area in the arctic. In this case, the evaporation of the oceans becomes less intense as the salinity increases, turning the atmosphere drier (see Figure 3), and consequently allowing less moisture for the growth of land glaciers.

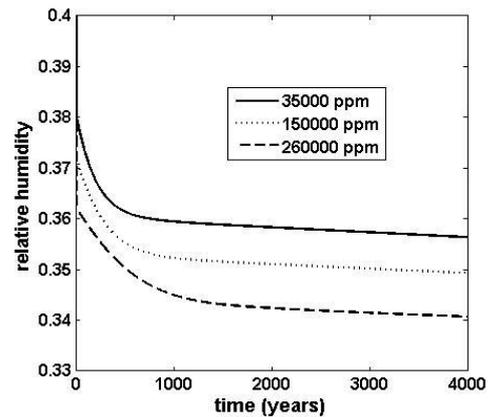

Figure 3 – Evolution of the global relative humidity for the case $\theta_0 = 0.7$, average ocean depth 3.7 km, and 3 different ocean salinities

For the scenario of $\theta_0 = 0.4$, the time-scale required for sea ice to fill the oceans of the arctic regions suffered a reduction (Figure 4), due to the reduction of ocean area; however, the relative difference between the salinity of 35,000 ppm and 260,000 ppm remained on the order of 100%. As for the time-scale for ice sheets to cover the land area, there was an increase (Figure 5), not only due to the increase of land area, but also because the less extensive oceans provided less water vapor to the atmosphere through evaporation, causing the average relative humidity to be less than the case of $\theta_0 = 0.7$ (Figure 6). Also, the relative difference of timescale between 35,000 ppm and 260,000 ppm remained on the order of 23%. Figures 4, 5 and 6 show the results.

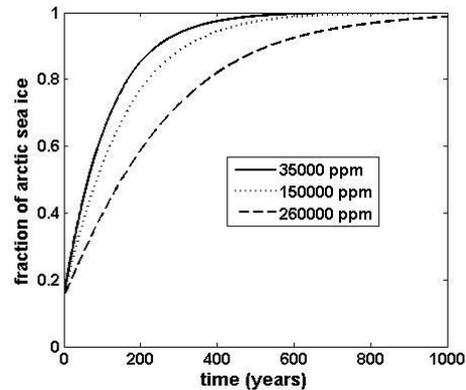

Figure 4 – Evolution of the fraction of arctic ocean area covered by ice, for the case $\theta_0 = 0.4$, average ocean depth 3.7 km, and 3 different ocean salinities.

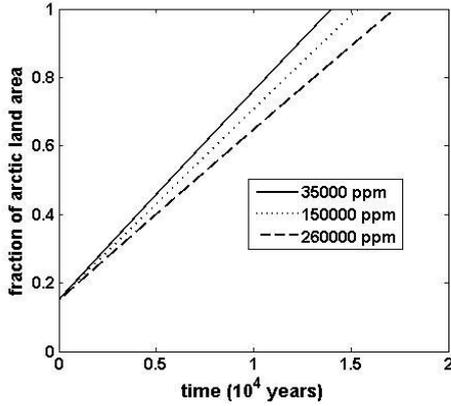

Figure 5 – Evolution of the fraction of arctic land area covered by ice, for the case $\theta_0 = 0.4$, average ocean depth 3.7 km, and 3 different ocean salinities.

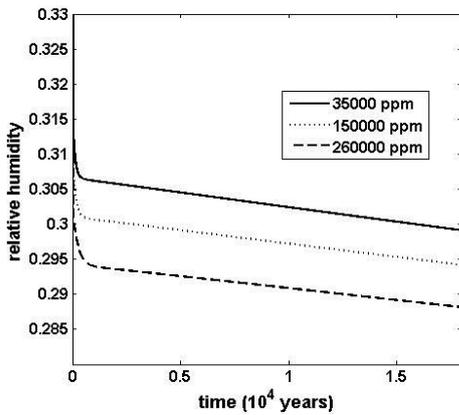

Figure 6 – Evolution of the global relative humidity, for the case $\theta_0 = 0.4$, average ocean depth 3.7 km, and 3 different ocean salinities.

Finally, in the most extreme scenario of $\theta_0 = 0.1$, the time required to fill the entire arctic land with ice glaciers ranges from 115,000 years (salinity = 35,000 ppm) to 145,000 years (salinity = 260,000 ppm), that is, an increase of around 26% between the maximum and minimum values of salinity. The small ocean area becomes covered in a much shorter time, as expected, but the difference in timescale between oceans with 35,000 ppm and with 260,000 ppm remains around 100%. Figures 7 and 8 show the results. This scenario represents a more intense effect of the salinity-evaporation correlation, since the ocean area is smaller and the evaporation rate is proportional to the liquid area exposed to the atmosphere. Consequently, the average atmospheric humidity is lower than the previous cases (Figure 9). The slight increase in the relative time difference (between oceans with 35000 ppm and 260000 ppm) in land glaciers (26%) compared with the cases of $\theta_0 = 0.7$ and $\theta_0 = 0.4$ (~23%) involves the action of a new phenomenon, which will be explored in the next section.

It is interesting to note that, for the same salinity, the necessary time to fill the arctic land area with ice increases considerably with diminishing ocean area, reflecting the fact that there is less available surface ocean area (proportionally) for evaporation, directly affecting the hydrological cycle.

Eq. (7) allows us to work regardless of the total atmospheric pressure at the surface of the planet, since $\phi$ is a dependent variable on the water mass balance, and $p^*$ depends only on the surface temperature. The value of $\alpha$, on its turn, does depend on the total atmospheric pressure, among other variables not considered in this simulation. For planets with total atmospheric pressure lower than the Earth's, $\alpha$ will increase, leading to a decrease of the time required to fill the arctic land area with ice sheets ($t_{fill}$).

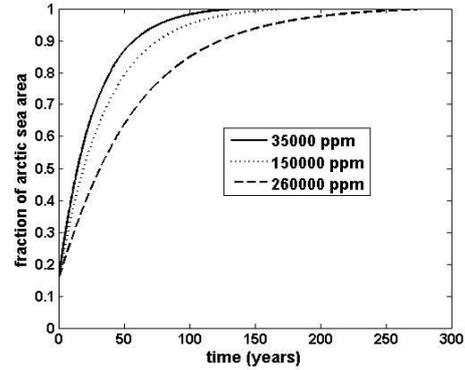

Figure 7 – evolution of the fraction of arctic ocean area covered by ice, for the case $\theta_0 = 0.1$, average ocean depth 3.7 km, and 3 different ocean salinities

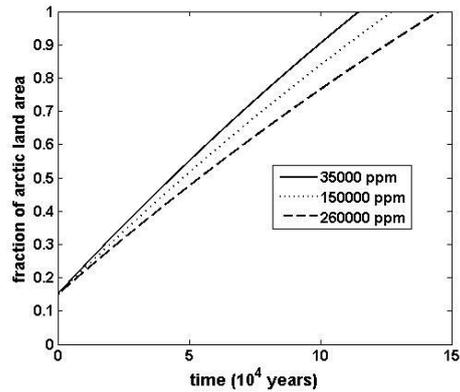

Figure 8 – evolution of the fraction of arctic land area covered by ice, for the case $\theta_0 = 0.1$, average ocean depth 3.7 km, and 3 different ocean salinities

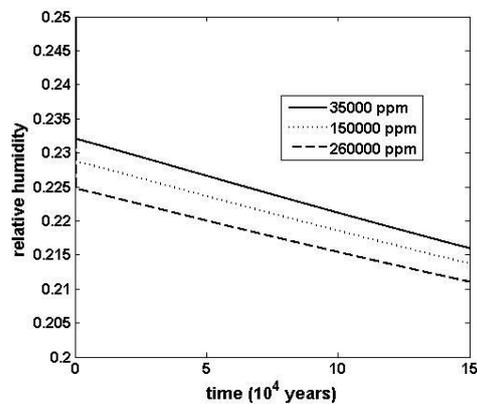

Figure 9 – Evolution of the global relative humidity, for the case $\theta_0 = 0.1$, average ocean depth 3.7 km, and 3 different ocean salinities.

However, our sensitivity analysis shows that, due to the non-linear nature of the differential-algebraic equations of the simulation, the impact of $\Delta\alpha$ (%) on $t_{fill}$ is small, that is, $|\Delta t_{fill}(\%)/\Delta\alpha(\%)| < 1$ for $\Delta\alpha$ of up to 50%. So, one must bear in mind that the values of $t_{fill}$ showed in figures 2, 5 and 8 may change slightly, depending on the total atmospheric pressure considered. Anyway, the effect of the salinity-evaporation phenomenon on the simulations is preserved, that is, $\Delta t_{fill}(\%)/\Delta salinity(\%)$ is a constant, for any value of $\alpha$.

## 6.2 Shallow oceans

In the simulations performed in the previous subsection, the salinity is also a function of time, since the salt mass in the oceans is constant in the timescales considered, and the ocean water mass is depleted along time. However, with an average depth of 3.7 km for the previous cases, the volume variation of the temperate and tropical oceans is not enough to increase their salinity to the point of affecting the water vapor saturation pressure, since they are linked with a non-linear relation (see Eq. (8)). Table 7 shows the details of volume and salinity variations for these simulations: both ocean volume and salinity show but small variations, even for the $\theta_0 = 0.1$ scenario.

|   | $\theta_0 = 0.7$ | | $\theta_0 = 0.4$ | | $\theta_0 = 0.1$ | |
|---|---|---|---|---|---|---|
|   | t=0 | t=3,250 | t=0 | t=14,000 | t=0 | t=115,000 |
| 1 | 6.865 | 6.833 | 2.207 | 2.171 | 0.245 | 0.220 |
| 2 | 5.253 | 5.186 | 1.688 | 1.613 | 0.188 | 0.139 |
| 3 | 35000 | 35,154 | 35,000 | 35,550 | 35,000 | 38,829 |
| 4 | 35000 | 35,433 | 35,000 | 36,575 | 35,000 | 46,793 |

Table 7 - Evolution of temperate (1) and tropical (2) ocean volumes ($10^8$ km$^3$) and their respective (3 and 4) salinities (ppm) for the three initial ocean area fraction scenarios (first row), and initial average ocean depth of 3.7 km, during the necessary time (years, second row) for the land ice glaciers to occupy all the available arctic land area.

The results change considerably (see Table 8) if we consider instead planets with shallow oceans, that is, those with average ocean depth on the order of hundreds of meters. By choosing 0.5 km as the initial value of average ocean depth, the time required for ice to cover the entire arctic region (the dynamics is controlled by the land ice expansion, even in the $\theta_0 = 0.7$ scenario) is increased substantially for $\theta_0 = 0.1$ (more than 300% increase) with a distinct non-linear relation with $\theta_0$, as the much more modest increase of 20% for $\theta_0 = 0.4$ shows. This phenomenon is caused by two combined and coupled phenomena. First, there is the more intense decrease of liquid water surface available for evaporation, since for the same area of ice glaciers created there will be a proportionally higher need of ocean water volume, and the surface area of the oceans (which affects directly the evaporation rate) is related to the variation of volume by an exponential. The second phenomenon, which develops simultaneously with the first, is the substantial salinity increase through time as ocean water evaporates, which increases the salinity-evaporation effect during the evolution of the system. This combined phenomenon is known as a dynamic negative feedback (on glaciation), that is, the salinity-evaporation effect resists the system evolution toward glaciation (through the initial salt content), and more strongly as the system moves away from the initial state.

|   | $\theta_0 = 0.7$ | | $\theta_0 = 0.4$ | | $\theta_0 = 0.1$ | |
|---|---|---|---|---|---|---|
|   | t=0 | t=3,450 | t=0 | t=16,800 | t=0 | t=500,000 |
| 1 | 9.276 | 8.952 | 2.982 | 2.589 | 0.331 | 0.074 |
| 2 | 7.098 | 6.436 | 2.282 | 1.566 | 0.254 | 0.040 |
| 3 | 35,000 | 36,221 | 35,000 | 40,100 | 35,000 | 140,570 |
| 4 | 35,000 | 38,464 | 35,000 | 50,176 | 35,000 | 187,470 |

Table 8 - Evolution of temperate (1) and tropical (2) ocean volumes ($10^7$ km$^3$) and their respective (3 and 4) salinities (ppm) for the three initial ocean area fraction scenarios (first row), and initial average ocean depth of 500 m, during the necessary time (years, second row) for the land ice glaciers to occupy all the available arctic land area, except for the case $\theta_0 = 0.1$, where a stable coverage of 70% is achieved.

The increase in time is non-linear with $\theta_0$, and for the case of $\theta_0 = 0.1$ the ice glaciers reach a stable extension of around 70% of the arctic land area (see Figure 10), although this does not mean that glaciation is effectively stopped, since the remaining surface area of temperate and tropical oceans are around 3% and 0.5% respectively (see Table 9).

Figure 10 shows the evolution of the arctic land area covered by ice glaciers for the cases of $\theta_0 = 0.1$, $\theta_0 = 0.2$ and $\theta_0 = 0.3$. In timescales approaching $10^5$ years or slightly longer, the feedback is strong enough to sharply curtail the rate of land ice formation. This effect will act simultaneously with periodic changes in planetary orbital elements (Croll-Milankovitch-like), which are known to act over comparable (20 to 40 x $10^4$ years) time-scales (Muller & MacDonald 2000), and that pushed the planet to an ice age in the first place. Given the timescales involved, our numbers suggest that this negative feedback might delay or even prevent the onset of ice ages and preserve most of the oceans from extensive water loss, and, perhaps, total freezing.

Such periodic changes of orbital elements, in the style of the Croll-Milankovitch cycles of the Earth, are probably common in terrestrial exoplanets, given the available sample of the architecture of multiplanetary systems revealed by exoplanet research. These changes arise essentially from mutual gravitational perturbations between planets, and in exoplanetary systems, if anything, they could well be on average more severe than observed in the Solar System, not only on Earth but also on Mars (Laskar et al. 2004).

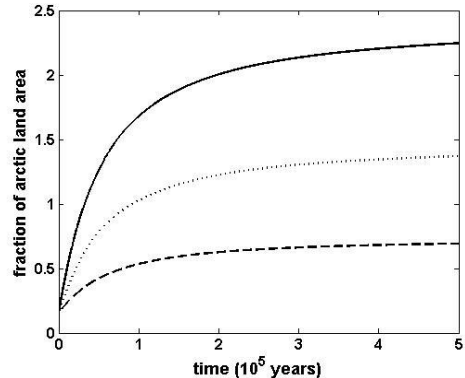

Figure 10 - Evolution of the arctic land area covered by ice glaciers for the cases of $\theta_0 = 0.1$ (dashed line), $\theta_0 = 0.2$ (dotted line) and $\theta_0 = 0.3$ (solid line). Values above 1 mean that the excess ice sheets occupy the temperate region.

These perturbations could well be relevant to climate evolution in low mass exoplanets given the high occurrence of planetary orbits locked in resonances, as well as the higher incidence of sizable eccentricities. Considering the timescales we found in this work, this effect would probably not be very important for Mars, since the periodicity of Martian ice-age episodes remains poorly known, controversial, and in the range of 100,000 to 400,000 years (Smith et al. 2016; Schorghofer 2016; Weiss 2019).

The negative feedback effect weakens considerably as the initial ocean coverage $\theta_0$ increases. This is explained by two facts, as $\theta_0$ increases: firstly, there will be less arctic land area to be filled with ice glaciers; and secondly, oceans will have larger water volume and feel less the water demand. Therefore, the volume drop in the oceans will be proportionally more modest as compared to that for $\theta_0 = 0.1$. This relation is also translated as a more modest increase in ocean salinities.

Also, for planets with higher values of $\theta_0$, the steady-state area of land ice sheets becomes more extensive. Although our model was not built for land ice sheets penetrating the temperate region (as we know they did during severe glaciation periods on the Earth), translated by values higher than one for the fraction of arctic land area covered by ice, we deem it robust enough for small extrapolations. Table 9 shows the results for $\theta_0$ in the range of 0.1 to 0.3, and the same time lapse of 5 x $10^5$ years. They indicate that the negative feedback could help delay or stop glaciation for a good range of $\theta_0$, when acting simultaneously with changes in orbital elements (Croll-Milankovitch-like).

| $\theta_0$ | Final fraction of arctic land area covered by ice | Final fraction of planet area covered by ice | Area of remaining oceans relative to the original ones (%) | |
|---|---|---|---|---|
| | | | Tropical | Temperate |
| 0.10 | 0.7 | 0.0519 | 0.5 | 3.0 |
| 0.15 | 1.0 | 0.0701 | 0.5 | 3.0 |
| 0.20 | 1.4 | 0.0924 | 0.5 | 3.0 |
| 0.25 | 1.8 | 0.1114 | 0.5 | 3.0 |
| 0.30 | 2.25 | 0.1298 | 0.5 | 3.0 |

Table 9 – Final extension of land ice sheets and remaining ocean areas for several values of $\theta_0$ and a time lapse of 500,000 years. Values above 1 for the fraction of arctic land area covered by ice mean that the excess ice sheets occupy the temperate region.

**6.3 Case study: Mars**

The previous simulations were designed for general exoplanets with a range of liquid water coverage in their surface. Moreover, the water coverage was considered the same for each of the three climatic regions. In the case of Mars, we have enough data on the hypothetical ancient ocean that covered the lowlands of the northern hemisphere (Saunders & Schneeberger 1989; Di Achille and Hynek 2010; Parker, Clifford & Parker 2001; Citron, Manga & Hemingway 2019) so that we can afford a customized simulation. Although there is still an ongoing debate as to whether or not the many signs on the surface of the planet are sure imprints of an ocean, or even if the planet ever had enough greenhouse effect to allow liquid water (Ramirez & Craddock 2018), we will assume that it existed in fact. The purpose of the simulation of the putative ocean on Mars is to assess how the salinity-evaporation effect could have delayed the onset of ice ages.

The topography of Mars shows that most of the low terrain is located in the northern hemisphere (with the exception of the Hellas Planitia and Argyre Planitia in the southern hemisphere, whose surface lies well below the topographic datum of Mars); The putative ancient ocean would have covered the entire northern arctic region, and sections of the northern temperate and tropical regions. The onset of an ice age would partly freeze the waters of the northern arctic and form glaciers in the highlands of the southern arctic regions through the transport of water vapor in the atmosphere. These highland glaciers would act as cold traps: evidence for their existence exists (Head & Pratt 2001) at the Dorsa Argentea region (mid Hesperian) and related units. The surroundings of the southern Argyre and Hellas basins also suggest glacial landforms with accumulation of past ice (Kargel & Strom 1992), as do some northern, high-altitude, near equatorial Hesperian landforms such as Ceraunius Tholus (Fasset & Head 2007).

There remains considerable controversy whether Mars´ early climate was capable of sustaining both northern oceans and southern, highland glaciers (Wordsworth 2016). Most recent analysis favor a cold and relatively dry Martian climate, yet present knowledge does not completely rule out northern oceans, either in an episodic fashion or as a transitional state between more stable end states. It is also important to keep in mind that, on Mars, runaway glaciation has high-altitude regions playing a key role as cold traps, by holding ice and snow due to atmospheric transport of water (Wordsworth 2016). On the Earth, the freezing of a near global ocean is the dominating event.

Taking into account that our model does not distinguish the northern from the southern regions (the equinox configuration dictates that both the northern and southern regions receive, on average, the same flux of stellar radiation), the reality of the planet is translated to the model as 50% water coverage for the (entire) arctic region (the average between 100% in the northern polar region and 0% for the southern one); this numerical manipulation is necessary since the simulation does not distinguish between the northern and southern hemispheres, considering only three climatic regions.

Following the same reasoning for the temperate and tropical regions, and considering the estimated water extent of the ancient northern ocean and of water bodies in Hellas and Argyre basins, we estimated a water extent of 35% and 30% for the temperate and tropical regions respectively. This is equivalent to an overall surface water fraction of 34.25%, very similar to the estimate of 35.7% made by Di Achille and Hynek (2010), and a more conservative one.

As for the total water volume, Di Achille and Hynek (2010) have estimated a value of 1.24 x 10$^8$ km$^3$, taking into account the northern ocean and the areas of Hellas and Argyre basins; however, they admit the possibility of overestimation due to the addition of ancient paleolakes in craters which might have been formed later. The volume of the Arabia ocean estimated by Citron, Manga & Hemingway (2019) is 5.5 x 10$^7$ km$^3$. So, we chose the intermediate value of 9 x 10$^7$ km$^3$ for our base case, which translates to an average water depth of 1.82 km, which is higher than our previous definition of shallow ocean, but lower than the average depth of 3.7 km used in the initial simulations.

The results of the simulations are shown in Figures 11 to 13 and in Table 10. As expected, the results for the arctic ocean are similar to the case of $\theta_0 = 0.7$ for general planets with deep oceans, since for the Martian arctic region we assumed $\theta_0 = 0.5$. As for the evolution of the land glaciers, the result is intermediate between the results for $\theta_0 = 0.7$ and $\theta_0 = 0.4$ for deep oceans, indicating an increase of ~23% in the time required to fill the arctic region, and considering the usual salinity range of 35,000 to 260,000. Table 10 shows that although the initial average depth of 1.82 km implies a stronger negative feedback due to increased salinity and decreased ocean area, when compared with the average depth of 3.7 km, it is still not strong enough to decrease the growth rate of land ice sheets considerably over time. Considering the much longer periodicity of Martian ice ages (~10$^5$ to a few times 10$^5$ years) as compared to that of terrestrial ice ages (about 40,000 years for the obliquity cycle), it is unlikely that the negative feedback due to salinity in putative Martian oceans could have made a difference. Known climatological cycles on Mars have much longer periodicity and are presumably stronger. A sizable influence could be obtained, however, if the average depth of such putative oceans was much less than the estimates used for the model. This possibility will be explored in future work.

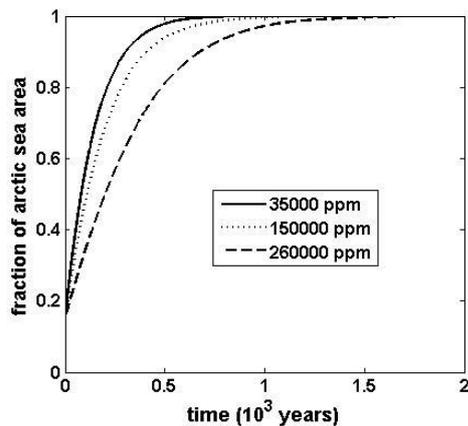

Figure 11 - Evolution of the fraction of Martian arctic ocean area covered by ice, for 3 different ocean salinities.

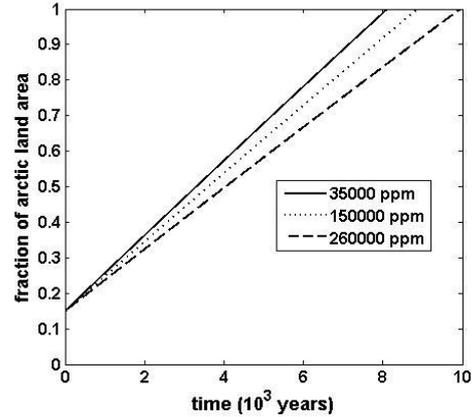

Figure 12 - Evolution of the fraction of Martian arctic land area covered by ice, for 3 different ocean salinities.

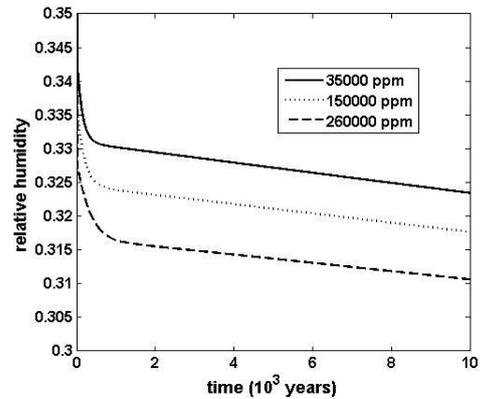

Figure 13 - Evolution of Martian global relative humidity, for 3 different ocean salinities.

We maintained our assumption of an ocean with mainly NaCl salt with three different initial concentrations. Mars apparently did not have a plate tectonics process like the one which exists in our planet (Nicola & Padovan 2019) and probably no hydrothermal systems as well, so its putative ancient ocean, only accumulated salt due to weathering of dry land; therefore, the salt content of the ocean was probably high during most of its existence. The discovery of sulfate deposits on Mars (Klingelhöfer et al. 2004, Rieder et al. 2004) leads to the conjecture that its ancient ocean and lakes might have been rich in sulfate salts, perhaps due to the interaction between the water and a sulfur-rich atmosphere (SO$_2$, SO$_3$). However, the average solubility of sodium and calcium sulfate in water at the temperatures of interest (0 to ~30 ºC) is well below that of sodium chloride (Perry 1997 b). Sulfide and sulfite salts have even lower solubilities. For calcium sulfate, the solubility ranges from 0.21 to 0.24 g per 100 ml of water, depending on the hydration degree, and for sodium sulfate the value ranges from 19 to 44 g per 100 ml of water at 20 C.

|  | Initial salinity 35,000 ppm | Initial salinity 150,000 ppm | Initial salinity 260,000 ppm |
|---|---|---|---|
| Final Vol. (temp.) | 4.039 | 4.054 | 4.072 |
| Final Vol. (trop.) | 2.459 | 2.485 | 2.515 |
| Final sal.(temp.) | 36,518 | 155,210 | 266,970 |
| Final sal.(trop.) | 39,228 | 164,280 | 27,890 |
| Initial Vol. (temp.) | 4.221 | | |
| Initial Vol. (trop.) | 2.769 | | |

Table 10 – Evolution of salinities and ocean volumes (10$^7$ km$^3$) for the three different initial salinities for the Mars simulations.

Further, since the molecular weight of sulfite and sulfate salts are much higher than that of NaCl, their molality, which drives the colligative effects under study in this work, would be very small. As for magnesium sulfate, which might have been a salt dissolved in Mars' waters, the solubility even surpasses the one for NaCl (Perry 1997 b), leaving only the molecular weight as a hindrance. The hypothesis of sulfate salt dissolved in water on Mars presents other complications. For example, if the sulfate originated from an atmosphere rich in $SO_2$ and $SO_3$, its massive dissolution in water would have altered its pH, consequently altering the solubility of salts.

It is not our intention to exhaust the ongoing debate about the complex sulfate problem on Mars (King, Lescinsky & Nesbitt 2004; Chow & Seal II 2007; Marion et al. 2013; Kaplan et al. 2016), or even suggest a new hypothesis. Our simulations involved only NaCl, based on the indication that the mantle of Mars is rich in Na and Cl, and poor in S (Yoshizaki & McDonough 2020). In what follows we provide the reader with a colligative effect equivalence, that is, for the initial value of NaCl concentration in ppm, we provide the equivalent value in ppm of $MgSO_4$, using the colligative equivalence equation

$$(iM)_1 = (iM)_2 \quad (16)$$

For the units used in this work (ppm of solute), this equivalence can be written as

$$\frac{i_1}{MW_1}\left[\frac{ppm_1}{1000 - \frac{ppm_1}{1000}}\right] = \frac{i_2}{MW_2}\left[\frac{ppm_2}{1000 - \frac{ppm_2}{1000}}\right] \quad (17)$$

where the subscript 1 refers to NaCl and the subscript 2 refers to $MgSO_4$. Table 11 shows the equivalent concentration of $MgSO_4$ for the initial concentrations of NaCl used in the simulations.

| Initial Conc. of NaCl (ppm) | Equivalent Conc. of $MgSO_4$ (ppm) | Observation |
|---|---|---|
| 35000 | 80663 | |
| 150000 | | Beyond $MgSO_4$ solubility |
| 260000 | | Beyond $MgSO_4$ solubility |

Table 11 – Conversion of concentrations between NaCl and magnesium sulfate for the same colligative effect.

The interested reader may substitute NaCl in the previous simulations with other salts by means of the equivalence equation, only taking care to check that the resulting concentration **does not surpass the value** of its solubility in water.

**6.4 Astrobiological implications**

Planets smaller than the Earth are expected to have high salt contents in their oceans, due to the faster planetary cooling process, which would have slowed down or even stopped plate tectonics and its associated hydrothermal system. These intrinsically more saline (deep) oceans will resist the glaciation process by slowing down the growth rate of land ice sheets and glaciers. The time extension could be up to around 23%, considering the maximum salinity range studied, between Earth's and the near saturation salinity. This effect could possibly counteract Croll-Milankovitch-like orbital forcings that push the planetary climate system to ice ages. From the gaian point of view (Lenton 1998; Nicholson et al. 2018), this stabilizing phenomenon can be viewed as an abiotic mechanism to avoid, or at least mitigate, ice ages potentially disastrous to life, since smaller planets would have less massive oceans, with less heat capacitance, and therefore more susceptible to total freezing (snowball events) than the oceans of larger planets.

At this juncture it is interesting to note that Snowball Earth global glaciation events have been linked to rapid oxygenation of Earth´s atmosphere (Hoffman 2018, Brocks, Jarrett & Silantoine et al. 2015, Vincent, Gibson & Pienitz el al. 2000), however, not without controversy (Warke, Di Rocco & Zerkle et al. 2020, Sperling, Wolock & Morgan et al. 2015). The last episode of global glaciation about 635 Myr ago has even been considered as one of the possible sparking causes of the Cambrian Explosion of marine life (Hoffman 2018): both the oxygenation and the spectacular Cambrian radiation of animal life are considered essential to the continuing flourishing of life on Earth.

While we acknowledge possible beneficial effects from global glaciations towards the biosphere, the proposed mechanisms whereby Earth eventually got rid of global glaciations involve its being a dynamic, large rocky planet, sporting continuing volcanism and plate tectonics. Also, despite a cold and solid surface, Snowball Earth had a fully dynamic atmosphere, cryosphere, ocean system and hydrological cycle (Hoffman 2018, Pierrehumbert 2005). For the smaller, rocky planets we focus on the present work, plate tectonics would be weak or nonexistent, volcanism is expected to peter out as the planet evolves, and the ocean system would be considerably less dynamic. In such smallish rocky worlds, for which shallow and highly saline oceans are the most likely to occur, global glaciation could be much more dangerous to life, owing to a much more limited suite of geophysical processes that might dynamically prompt negative feedback to reverse Snowball Earth events.

The appearance of life depends upon many physical factors (Lingam & Loeb 2019), not to mention biochemical factors, but it is thought that liquid water is a necessary requirement, as the definition of the Habitable Zone attests. Therefore, if we assume that life in other planets is born preferentially in oceans and spends most of its existence there, before migrating to dry land, as has happened on Earth, the salinity-evaporation effect could be viewed as an extra protection mechanism.

Planets with shallow oceans, those with the highest susceptibility to disastrous ice ages, benefit from the strengthening of the salinity-evaporation effect with time. The process of glaciation, acting simultaneously with the reduction of ocean surface area in a dynamic negative feedback, reinforces the opposition to other forcings, mitigating and even effectively stopping glaciation. Besides the direct danger to life mentioned above, intense ice ages can also have other deleterious consequences to life productivity, like an excessive reduction of ocean coverage fraction (Lingam & Loeb 2019 b).

The marine life of such oceans would have to develop adaptation strategies to withstand high salinity variations on time scales of $10^4 - 10^5$ years. This is, in principle, a viable evolutionary scenario, since on Earth there are extremophiles which thrive in high-concentration brines, besides ocean life adapted to normal salinity. Indeed, Cubillos et al. (2018) report both Archaea and Bacteria genera thriving at moderate altitude, in the Chilean Atacama desert, in lakes presenting up to 35% salinity under natural conditions, and even 56% salinity (350,000 and 560,000 ppm respectively) for industrially enhanced mining waste. The latter result pertains to LiCl, which is interesting since it shows adaptability extending beyond NaCl, a far more commonly studied salt, additionally suggesting that microorganisms can quickly and successfully adapt to changing salinity conditions.

Once the glaciation process is reverted, due to a number of possible factors, the return of land ice to the oceans as liquid water will restore a lower level of salinity, and a more extensive ocean area, which is able to evaporate more easily, increasing the greenhouse effect that had already been increased in order to end the ice age. That is, the salinity-evaporation effect, in shallow oceans, is also a destabilizing (positive feedback) agent regarding planet warming. But, considering also that smaller planets are more susceptible to cooling, due to a probably thinner atmosphere, this destabilizing factor may also turn out to be key to life survival, for the likelier planetary habitability bottleneck would tend to be excessively cold climates, rather than hot ones. If so, the salinity-evaporation effect could be viewed as an abiotic climatic self-regulation, always aimed at keeping the planet away from excessive glaciation, although the name "cycle" would possibly not be a proper one, as contrasted to the carbonate-silicate cycle (Kasting 2010; Rushby 2018), which acts as a negative feedback for both high and low temperatures, .

It is interesting to note that, as long as greenhouse effect goes, the salinity-evaporation effect also acts as a destabilizing (positive feedback) agent on glaciation, for the lower drier atmosphere, caused by the depleted evaporation, helps the decrease of surface temperatures. Although our model is not rigorous with respect to the greenhouse effect caused by water vapor, we did take it into account, and we think that the salinity-evaporation effect (negative feedback) on the growth rate of ice glaciers outweighs the humidity factor, either by intrinsically high salinity or by the coupling of high salinity with the feedback of shallow oceans. However, more complete simulations are needed to assess more precisely the relative forces of these agents, which we intend to pursue in a forthcoming paper

Possibly, the increased retention of liquid water and lower water vapor pressure provided by the salinity-evaporation relationship could help delay the volatile loss to space in planets subject to high EUV/X-rays flux from the parent star. This phenomenon is not explicitly approached by our model, which moreover deals with timescales of $10^2$-$10^5$ years, while volatile loss is a process with timescales of $10^8$-$10^9$ years. Yet the influence of salinity on volatile loss would be a logical consequence, since a drier atmosphere would convey a smaller amount of water vapor to the stratosphere, where molecules are split by radiation and particle fluxes, and lost to space. Quantitative studies should be performed to assess this possibility, and it is our intention to do so in a forthcoming paper.

**6.5 Toy Model versus Global Circulation Models**

Our simple model was developed to check if an oceanic salinity-evaporation negative feedback could limit the growth of land glaciers in rocky habitable planets, under general, easily controlled yet physically realistic, conditions. While Global Circulation Models (GCM) are richer in physical interactions that affect the climate, and we hope they will be used in the future to probe particular types of habitable planets, our simple model is a more appropriate first approach, and we now list some arguments to justify our assumption.

One of the more important advantages of GCM models – probably the most important one – is the incorporation of ocean circulation, since liquid water has a considerable heat capacity and its currents help to distribute heat across a wide range of latitudes (Charnay et al. 2013; Gómez-Leal, Codron & Selsis 2016; Paradise and Menou 2017; del Genio et al. 2019; Olson et al. 2020). For oceans with high salinity the circulation pattern may even differ from the usual outcome of simulations (Cullum, Stevens & Joshi 2016). Even in GCM models where the atmosphere circulates, but oceans do not (Edson et al. 2011), a horizontal heat diffusion for oceans is included to account for heat transport.

However, the scenarios we are working with, namely low global ocean coverage (except for the 70% case, which is essentially a control case and shows but a weak salinity effect), limit ocean circulation in latitude, since it is more probable that this coverage is isolated in fragmented bodies of water. Even if we consider a sole water body for the 10% and 40% cases, and low water coverage in general, it is likely that such body will have a limited latitude span. We have thus assumed equal water coverage for each climatic region, and a GCM model is unlikely to improve the results of our simulations significantly.

The above line of reasoning leads directly to the next, that is, the finer surface spatial resolution provided by GCM models, compared to the crude division of 3 regions developed in our toy model. Since we are working with hypothetical habitable worlds with no specific mapping of land and oceans, the advantages of finer resolution are negated. Therefore, we believe the simplicity of the toy model is an advantage in terms of both generality and robustness for a first investigation.

Our assumption of constant cloud coverage, distributed equally between the climatic regions, is again a crude simplification, and GCM models might improve the results. However, cloud formation in Earth-like exoplanets is notoriously complex (Zsom, Kaltenegger & Goldblatt 2012, Charnay et al. 2013) even for models which strive to reproduce cloud cover on Earth during small periods of time (Probst et al. 2012). In exoplanets atmospheric composition changes and other factors make the problem even more complex. Instead we decided to mimic the effect of clouds on climate using only their albedo (Kasting, Pollack & Ackerman 1984).

The assumption of homogeneous humidity throughout the entire area of our model planets, which reacts instantly to variations in evaporation and condensation, is also a simplification worth of comment, when contrasted with the equations of air transport in GCM models. The distribution of humidity depends strongly on planetary topography, coupled with the land and ocean distribution, which are unknown parameters in our simulations for general planets. Also, for small planets, with wind speeds comparable to that of Earth's (Chittenden et al. 2018, Temel et al. 2019, Soria-Salinas et al. 2020), the homogenization process is faster, with a time scale much shorter than the effects of salinity on climate. This probably helps to explain the small interannual variation of water vapor on Mars (Smith et al. 2018), as well as its small variation in latitude, except in the polar regions. So, we think that a homogeneous humidity abstraction did not influence the main results in a significant way.

Our decision to use thinner surface atmospheric pressures as the model planet becomes smaller is based on hard evidence in our Solar System, but that the same pressure holds for the vertical profile of the troposphere (taken into account in GCM models) is not. This simplification affects the water holdup in the atmosphere, but, again, we believe that the results of the simulations will be affected only marginally, since the bulk of water vapor resides in the first few kilometers of the troposphere, where the pressure decrease is measured in percent points, rather than orders of magnitude. Also, our evaporation rate model does not depend on atmospheric pressure (translated as air density) as it does in GCM models (Charnay et al. 2013), which is a robust assumption for thin atmospheres.

In short, the precision offered by GCM models is not a significant advantage for our exploratory approach of small and dry habitable planets. Considering the many unknown parameters, and the robustness of the overall physical description translated into the equations of the toy model, we think a simple approach suffices to analyze the salt-evaporation effect with some detail. We acknowledge that simulations richer in specific parameters, including distinct atmospheric compositions (the presence or not of a layer of ozone is important; Gómez-Leal et al. 2019), and the formation of organic hazes from $CO_2$ and $CH_4$, would benefit directly from the use of GCM models. Other factors that would improve our analysis, along with GCM modelling, are: taking into account the greenhouse effect from other molecules other than water, a more detailed continental and oceanic distribution over planetary surface, and better precipitation models.

# 7 CONCLUSIONS

We present a planetary climate toy model, capturing the essentials of ocean evaporation and ice formation dynamics and coupled to radiative equilibrium temperatures. We apply it to case studies involving small rocky planets, on the verge of entering an ice age, with intrinsically high salinities and a low volatile content. The model seeks to study whether intrinsically high salinities affect the glaciation process; we also analyze negative feedback mechanisms thereby increased salinity, created by water evaporation in shallow oceans, can enhance the resistance to the growth of glaciers on land. The lowered evaporation rate created by an increased salinity, especially when coupled with the negative feedback, might thus help mitigate or even avert unchecked growth of land ice, saving the planet from a long spell of glaciation and its adverse effects on marginal biospheres living off drier and colder planets when compared to Earth's conditions.

Our main conclusions are:

1- our model clearly indicates that the salinity-evaporation correlation may act as a climatic buffer in planets entering ice ages through a considerable increase in the time necessary to fill arctic land with ice sheets, up to ~23% if we consider a wide salinity range, from 35,000 to 260,000 ppm;

2- for planets with shallow oceans (with average depth in the order of hundreds of meters), the self-reinforcement of the salinity effect, acting simultaneously with the reduction of ocean area available for evaporation, acts very strongly as a negative feedback on the glaciation process, to the point of sharply decreasing or even effectively stopping the land ice sheet growth rate during the time-scale of the simulations. This effect is a function of, mainly, the ocean salinity and the fraction of the planet area covered by oceans.

3- the longer timescale for the growth of land ice in planets with shallow and highly saline oceans is faster or comparable to the timescale of Croll-Milankovitch-like changes of planetary orbital elements, which are a fundamental mechanism to trigger ice ages in rocky planets. It is thus possible that the salinity-evaporation feedback effect on small planets with shallow oceans (or water-deficient, Earth-sized planets) may delay ice ages or even prevent them from developing altogether;

4- the assumed high salinities we employed are expected for habitable planets comparable in size to the Earth and smaller, for which the plate tectonics process is probably less intense and consequently hydrothermal systems operate less effectively in removing salt from oceans, provoking enhanced ocean salinity;

5 – we performed a customized case study for a putative past ocean on Mars, adapting the model to hypothesis concerning the ocean´s location and water volume. We conclude that in the case of Mars the salinity-evaporation effect lies midway between deep and shallow oceans, and moreover that the resulting timescales (~$10^3$ years) would be much faster than the periodicity of Martian ice ages (~$10^5$ years). Most probably this effect did not play a role in influencing Mars´ past climate;

6- such salinity-evaporation correlation in shallow, saline oceans, reinforced by the dynamic negative feedback in extreme cases, can be understood as an abiotic climatic self-regulation cycle, explored for the first time in the present work. This feedback helps keep small planets, possessing weak or no plate tectonics and which are deficient in volatiles, away from extremes of glaciation, which could otherwise freeze the oceans to the bottom and kill all marine life, since the heat capacitance of such oceans cannot match that of Earth-like planets, with vast and deep oceans;

7- the effect of salinity on evaporation could conceivably mitigate the loss of volatiles to space for small habitable planets, since their atmospheres tend to be drier and cooler, thus helping a planet maintain a dry stratosphere and avoiding high altitude photolysis by EUV radiation. This effect might be even more important for atmospheres that lack the temperature inversion which defines, on Earth, the troposphere-stratosphere boundary;

8- the salinity-evaporation effect, plus its reinforcement as a negative feedback for shallow oceans, is probably a common phenomenon for rocky exoplanets in the Habitable Zone. We suggest that these effects be incorporated in future climatic models, and in astrobiological considerations regarding physical conditions that could enhance the resilience of life in the Universe;

**ACKNOWLEDGEMENTS**

We thank the anonymous referee for suggestions and criticism that considerably improved the manuscript. Rafael Pinotti would like to thank André Domingues Quelhas for his help with Matlab technical details during the design of the simulation environment. GFPM acknowledges financial support by the Conselho Nacional de Desenvolvimento Científico e Tecnológico (CNPq/Brazil). We would like to thank Ignasi Ribas, José Dias do Nascimento Jr., Vladimir Airapetian and Aline Vidotto for valuable discussions. The research has made use of the SIMBAD data base, operated at CDS, Strasbourg, France, and of NASA Astrophysics Data System.

**DATA AVAILABILITY**

The data underlying this article are available in the article and in its online supplementary material.